\documentclass[12pt]{article}
\usepackage{cite}
\usepackage{amsmath, amsthm, amssymb, graphicx,slashed}

\numberwithin{equation}{section}

\def\be{\begin{equation}}
\def\ee{\end{equation}}
\def\ba{\begin{align}}
\def\ea{\end{align}}
\def\beq{\begin{eqnarray}}
\def\eeq{\end{eqnarray}}
\def\a{\alpha}

\begin{document}

\begin{titlepage}

\vskip 1.5in
\begin{center}
{\bf\Large{Elliptic Genera of Non-compact Gepner Models}\vskip0cm
\bf\Large{and Mirror Symmetry} }\vskip 0.5cm
{Sujay K. Ashok$^a$ and Jan Troost$^b$} \vskip 0.3in 

\emph{$^{a}$Institute of Mathematical Sciences\\
  C.I.T Campus, Taramani\\
  Chennai, India 600113\\ }

\vspace{.1in}

 \emph{\\${}^{b}$ Laboratoire de Physique Th\'eorique}\footnote{Unit\'e Mixte du CNRS et
    de l'Ecole Normale Sup\'erieure associ\'ee \`a l'universit\'e Pierre et
    Marie Curie 6, UMR
    8549.
}\\
\emph{ Ecole Normale Sup\'erieure  \\
24 rue Lhomond \\ F--75231 Paris Cedex 05, France}
\end{center}
\vskip 0.5in

\baselineskip 16pt

\begin{abstract}

\vspace{.2in}
We consider tensor products of $N=2$ minimal
models and non-compact conformal field theories with $N=2$
superconformal symmetry, and their orbifolds.  The elliptic genera of
these models give rise to a large and interesting class of real Jacobi
forms.  The tensor product of conformal field theories leads to a
natural product on the space of completed mock modular forms.  We
exhibit families of non-compact mirror pairs of orbifold models with
$c=9$ and show explicitly the equality of elliptic genera,
including contributions from the long multiplet sector. 
The Liouville and cigar deformed elliptic genera transform into each
other under the mirror transformation.

\end{abstract}
\end{titlepage}
\vfill\eject

\tableofcontents

\section{Introduction}
The study of two-dimensional conformal field theories in terms of
their minimal model description, their Landau-Ginzburg phase or as
gauged linear sigma-models has proven to be very useful
\cite{Witten:1993jg, Witten:1993yc}.  It has taught us about the
space of two-dimensional conformal field theories, and its geometrical
structure. The study has had a profound impact on our understanding of
compact Calabi-Yau manifolds and mirror symmetry, and it has had
interesting applications in the field of singular manifolds and toric
varieties (see \cite{Hori:2003ic} for a review).

The extension of this study to include theories with non-compact
targets, and in particular non-compact Calabi-Yau manifolds is very
interesting. It is a natural generalization from the perspective of
studying Calabi-Yau manifolds locally, or from the viewpoint
of understanding holography in curved non-compact spaces that
asymptotically have a linear dilaton profile
\cite{Aharony:1998ub,Giveon:1999zm}. This field has already given rise
to many results including a study of the map between deformations of
the geometry and the spectrum of non-compact conformal field theories \cite{Eguchi:2000tc,
  Eguchi:2000cj, Eguchi:2004yi, Eguchi:2004ik,Ashok:2007ui}, mirror symmetry for
non-compact Gepner models, as well as an intriguing relation between
orbifolds in asymptotically linear dilaton spaces and flat space toric
orbifolds \cite{Ashok:2007ui}. However many of the results have been
based on studying the chiral (anti-chiral) rings of the theory.

Recently, there has been a lot of progress in our understanding of the
elliptic genus of non-compact $N=2$ superconformal field theories with
central charge larger than three \cite{Troost:2010ud, Eguchi:2010cb,
  Ashok:2011cy}.  In particular, it was understood that the elliptic
genus is modular covariant and real. Non-holomorphic contributions
arise from the continuous part of the spectrum of the
two-dimensional conformal field theory. This has led to a physical
understanding of the modular completion of mock modular forms in terms
of both a modular Lagrangian path integral description \cite{Troost:2010ud}, and a
Hamiltonian viewpoint in terms of an integral over a difference of
spectral densities for right-moving primary bosons and fermions \cite{Ashok:2011cy}.

In this paper we apply these new insights to the study of conformal field
theories which are tensor products of $N=2$ minimal models and $N=2$
Liouville theories (or $N=2$ cigar coset models), and their
orbifolds. The elliptic genus of tensor product theories is the
product of the individual elliptic genera. For orbifold theories, we
can often identify the elliptic genus via standard twisting
procedures. 

For orbifolds of products of compact Gepner models, there have been many
interesting results \cite{Kawai:1993jk, DiFrancesco:1993ty,
  Berglund:1993fj, Berglund:1994qk, Kawai:1994fm}, especially in the
context of mirror symmetry. Given a Landau-Ginzburg formulation of a
compact Calabi-Yau, there is an algorithmic way to construct the
mirror. Most of the results on elliptic genera rely on the fact that, given the Poincar\'e polynomial of the theory,
there is a unique extension to the elliptic genus and the
identification of mirror pairs thus becomes simpler.

For non-compact conformal field theories, our generic construction will
give rise to a large new class of real Jacobi forms.  In particular,
when more than one non-compact model is involved, the 
product of elliptic genera gives rise to a modular completion of
the product  of mock modular forms. 
 Conformal field theory
elliptic genera thus provide a natural way to complete the product of
two mock modular forms. 

We apply this general reasoning to non-compact Gepner models and their
orbifolds in type II string theory.  The knowledge we gained about
non-compact elliptic genera allows us to check mirror symmetry
explicitly in these models in a way coherent with modularity and
ellipticity. The check includes some long multiplet contributions. Our
analysis is constructive in the sense that, starting from the
elliptic genus of a given orbifold theory we rewrite it 
such that the final expression has a natural interpretation as
the elliptic genus of the mirror model. Under the mirror transform, the
Liouville factors naturally go over into their cigar counterparts.

The paper has the following organization.  In section \ref{genera} we
discuss the elliptic genera of the basic models, which are the minimal
models with $c<3$ and the two types of non-compact models with $c>3$,
the Liouville and cigar theories. We also describe how to put these
together and construct the elliptic genera of tensor product and
orbifold models.  In section \ref{mirrorsection} we restrict to
orbifold models which are non-compact generalizations of Gepner models
in type II string theory.  Examples with central charge $c=6$ are
provided in section \ref{c=6} and those with central charge $c=9$ are
discussed in section \ref{c=9}. The technical ingredients necessary
for the calculations in these sections are provided in appendix
\ref{characters}. We end in section \ref{mockmodular} with a number of
proposals for how to extend our class of examples to a broader domain.

\section{Elliptic genera}
\label{genera}
In this section, we review the elliptic genera of $N=2$ minimal models and $N=2$ superconformal field theories with central charge greater than three, since these conformal field theories form the building
blocks of the models we study in sections 
\ref{mirrorsection} and \ref{c=9}.
 We also pause to make a point
about an embryonic example of mirror symmetry.
\subsection{Definition and Properties}
We study $N=2$ superconformal field theories  with a left $U(1)_R$ charge
$J_0$ and a right charge $\bar{J}_0$, as well as scaling
dimension operators $L_0$ and $\bar{L}_0$. The elliptic
 genus $\chi$ \cite{Schellekens:1986yi, Witten:1986bf}
is defined as a twisted partition sum with periodic boundary conditions
for the fermions:
\begin{eqnarray}
\chi(q,z) &=& 
\mbox{Tr} \, \, (-1)^F
q^{L_0-\frac{c}{24}} 
\bar{q}^{\tilde{L}_0-\frac{c}{24}} z^{J_0}.
\end{eqnarray}
We will also use the notation $\chi(q,z) \equiv \chi(\tau,\alpha)$ for
the elliptic genus, where the arguments are related through the
equations $q = e^{2 \pi i \tau}$ and $z = e^{2 \pi i \alpha}$. The
elliptic genus has elliptic and modular covariance properties which
make it 
a Jacobi form.
\subsection{The building blocks}
In this subsection, we list the elliptic genera of the elementary
building blocks that we will use to construct our models.
\subsubsection{The $N=2$ minimal models}
The elliptic genus of an $N=2$ minimal model with central 
charge $c=3- \frac{6}{k}$ and $k$ a positive integer is given by \cite{Witten:1993jg}:
\begin{eqnarray}
\chi(k;-)(q,z) &=& 
\frac{\theta_{11} (q, z^\frac{k-1}{k} )}{\theta_{11} ( q, z^\frac{1}{k} )}.
\end{eqnarray}
This is also the elliptic genus of the compact Landau-Ginzburg model 
with superpotential $W = X^k \,.$
It can, moreover, be derived from the gauged Wess-Zumino-Witten description of the model. We denote the level of the minimal model as an extra argument for the elliptic genus, followed by a semicolon. The elliptic genus of the minimal model has an expansion in terms of twisted Ramond sector characters \cite{Witten:1993jg, DiFrancesco:1993dg}:
\be
\chi(k;-)(q,z) = \sum_{j=0,\frac{1}{2},\ldots}^{\frac{k-2}{2}} {\cal C}^{j}_{2j+1}(q,z) \,.
\ee
The basic definitions and the modular and elliptic properties of these
characters are reviewed in appendix \ref{characters}.

\subsubsection{The $N=2$ Liouville model}
Next, we consider models with central charge $c=3+\frac{6}{l}$ with
the level $l$ equal to a positive integer.
The $\mathbb{Z}_l \subset U(1)_R$ orbifold of the $SL(2,\mathbb{R})_l/U(1)$
coset theory with central charge $c=3+\frac{6}{l}$ has elliptic
genus \cite{Troost:2010ud}:
 \be
 \chi(-;l)(q,z) 
 =  \frac{i \theta_{11}(q,z)}{\eta^3} \hat{A}_{2l} (z^{\frac{1}{l}},z^2  ; q).
\label{ncgenus}
 \ee
The level $l$ of the non-compact model follows the semicolon. This
 elliptic genus is also the genus of a generalized
 non-compact Landau-Ginzburg model 
with superpotential $W=e^{-lY}$, coinciding with
$N=2$ Liouville theory at radius $R=\sqrt{l \alpha'}$. 

Let us discuss these points in some detail, since it provides
an important embryonic example of mirror symmetry that pervades the rest
of our paper. Note that there are two known ways to obtain the
expression (\ref{ncgenus}) for the elliptic genus. The first way is
through the non-compact Landau-Ginzburg model, where one identifies
the R-charges of the fields, and their proper configuration space,
then to do a free field calculation to obtain the holomorphic part of
the elliptic genus \cite{Troost:2010ud}. A scattering calculation
using the Landau-Ginzburg potential will then further provide the
remainder term in the elliptic genus \cite{Ashok:2011cy}, thus proving
that expression (\ref{ncgenus}) is the elliptic genus of $N=2$
Liouville theory at radius $R=\sqrt{l \alpha'}$.
  Alternatively, a path
integral calculation shows that this is also the elliptic genus
of the $\mathbb{Z}_l$ orbifold of the cigar coset conformal field
theory \cite{Troost:2010ud, Ashok:2011cy}. This provides further
evidence for the equivalence of these models \cite{Hori:2001ax, Israel:2004jt} in
terms of the match of a modular covariant partition sum.

The mock modular form, of which the elliptic genus is the completion,
is a holomorphic Appell-Lerch sum which has an expansion in terms of
twisted Ramond $N=2$ superconformal characters $Ch$ extended by spectral flow
(see \cite{Israel:2004jt, Ashok:2005py} for our conventions for the arguments):
\be
\chi_{hol}(-;l)(q,z) = \sum_{2j-1=0}^{l-1}Ch(j;-\frac{1}{2}; q, z).
\ee
\subsubsection{The coset conformal field theory}
The $SL(2,\mathbb{R})_l/U(1)$ supersymmetric coset theory, which we refer to as the cigar theory, has an elliptic genus obtained by taking the $\mathbb{Z}_l$ orbifold of the elliptic genus quoted above. It is given by
\be
\chi(-;l)^{\mathbb{Z}_l}(q,z) = \frac{1}{l}
\frac{i \theta_{11}(q,z)}{\eta^3} \sum_{m_a,m_b\in  \mathbb{Z}_l} 
e^{- \frac{2 \pi i m_a m_b}{l}}
q^{-\frac{m_a^2}{l}} \hat{A}_{2l} (z^{\frac{1}{l}} q^{\frac{m_b}{l}}
e^{ \frac{ 2 \pi i m_b}{l}}, z^2 ; q).
\ee
We have denoted the orbifold group as a superscript to the elliptic
 genus. It is also the elliptic genus of $N=2$ Liouville theory at
radius $R= \sqrt{\alpha'/l}$. The holomorphic part of the elliptic genus can again
be expanded in terms of the extended characters:
\be
\chi_{hol}(-;l)^{\mathbb{Z}_l}(q,z) = \sum_{2j-1=0}^{l-1}Ch(j;-\frac{1}{2}-(2j-1); q, z)\,.
\ee

\subsection{Tensor product theories}
An elementary but important point is that the
 elliptic genus of a tensor product conformal field theory is the
product of the individual elliptic genera:
\begin{eqnarray}
\chi(\otimes_i CFT_i) &=& \prod_i \chi (CFT_i).
\label{product}
\end{eqnarray}
For example, for the tensor product of compact and non-compact
Landau-Ginzburg models with central charges associated to the positive
and integer levels $(k_1,k_2,\dots, k_p; l_1, l_2, \dots, l_q)$, the
elliptic genus reads:
\begin{eqnarray}
\chi (k_1,k_2,\dots, k_p; l_1, l_2, \dots,
l_q) (q,z) &=& 
\prod_{i=1}^p
\frac{\theta_{11} ( q,z^{1-\frac{1}{k_i}})}{\theta_{11} ( q, z^\frac{1}{k_i} )}
\prod_{j=1}^q
\frac{i \theta_{11}(q,z)}{\eta^3} \hat{A}_{2l_j} (z^{\frac{1}{l_j}},z^2  ; q).
\end{eqnarray}
One can generalize this elliptic genus
to one which keeps track of the R-charges of the individual factor
theories. We find a generalized elliptic genus:
\begin{align}
\chi (k_1,k_2,\dots, k_p; l_1, l_2, \dots,
l_q) (q,z_i,z_j) = 
\prod_{i=1}^p
\frac{\theta_{11} ( q,z_i^{1-\frac{1}{k_i}})}{\theta_{11} ( q, z_i^\frac{1}{k_i} )}
\prod_{j=1}^q
\frac{i \theta_{11}(q,z_j)}{\eta^3} \hat{A}_{2l_j} (z_j^{\frac{1}{l_j}},z_j^2  ; q)\,.
\end{align}
This is one of many generalizations of the twisted index. 
One can write down similar expressions where we replace some of the Liouville factors with cigar coset theories.

\subsection{Twisted blocks}\label{simporbs}\label{twistblocks}

In the following, we consider orbifolds of tensor products of the above models. 
For simplicity, we restrict our orbifold groups to be discrete
subgroups of the product of the $U(1)$ R-symmetry groups of the factor models.
In these circumstances, it is straightforward to generalize the
techniques of \cite{Kawai:1993jk} to describe the twisted partition sums from
which we build the elliptic genus of the orbifold.
In each factor theory, we have partition sums in the sectors twisted
by the generator of an orbifold group $\mathbb{Z}_n$ to the power $m_a
\in \mathbb{Z}_n$ and we can insert an operator corresponding to a
generator of the orbifold group to the power $m_b \in
\mathbb{Z}_n$. We then obtain the twisted partition functions:
\begin{eqnarray}\label{chimn}
\chi_{m_a,m_b} (q,z) &=& 
e^{2 \pi i \frac{c}{6} m_a m_b}
e^{2 \pi i \frac{c}{6} (m_a^2 \tau+ 2 m_a \alpha)}
\chi (\tau, \alpha+ m_a \tau + m_b).
\end{eqnarray}
The transformation properties of these twisted elliptic genera are
(with $\lambda,\mu \in \mathbb{Z}$):
\begin{eqnarray}
\chi_{m_a,m_b} (-\frac{1}{\tau},\frac{\alpha}{\tau}) &=&
e^{2 \pi i \frac{c}{6} \frac{\alpha^2}{\tau}}
\chi_{m_b,-m_a} (\tau, \alpha)
\nonumber \\
\chi_{m_a,m_b} (\tau+1,\alpha) &=&
\chi_{m_a+m_b,m_b} (\tau, \alpha)
\nonumber \\
\chi_{m_a,m_b} (\tau,\alpha+\lambda \tau + \mu)
&=& e^{2 \pi i \frac{c}{6} (m_a\mu-m_b\lambda-\lambda \mu)}
e^{- 2 \pi i \frac{c}{6} (\lambda^2 \tau + 2 \lambda \alpha)}
\chi_{m_a+\lambda,m_b+\mu} (\tau,z).
\end{eqnarray}
We assign a canonical phase factor to each factor model:
\begin{eqnarray}
\epsilon(m_a,m_b) &=& (-1)^{m_a+m_b+m_a m_b},
\end{eqnarray}
which will ensure that the total orbifolded model is free of discrete torsion.
For the partition sum including the phase, we use the notation:
\be
\tilde{\chi}_{m_a,m_b} = \epsilon(m_a,m_b) \chi_{m_a,m_b}.
\ee
The twisted building blocks for the R-symmetry orbifolds can be
simplified using the ellipticity and modular properties of theta
functions and completed Appell-Lerch sums $\hat{A}$ (see appendix
\ref{characters} for details).  It will be
convenient to express the twisted building blocks $\tilde{\chi}_{m_a,m_b}$
of the minimal models and the non-compact conformal field theories in terms of the twisted
Ramond sector characters of the conformal field theory. This renders
the transformation properties of each term under the insertion
of a generator in the trace manifest. 
We find the twisted blocks:

\begin{itemize}

\item for the minimal models
\begin{align}\label{chiMMchar}
\tilde{\chi}_{m_a,m_b}(k;-) &=e^{-\frac{2\pi i m_am_b}{k}}\sum_{j=0,\frac{1}{2},\ldots}^{\frac{k-2}{2}}
e^{\frac{2\pi i m_b}{k}(2j+1)} {\cal C}^{j}_{2j+1-2m_a}(q,z)\,.
\end{align}

\item for the anti-diagonal (or $\mathbb{Z}_k$ orbifolded) minimal models
\begin{align}\label{chiMMchar2}
\tilde{\chi}_{m_a,m_b}(k;-)^{\mathbb{Z}_k} &=
e^{-\frac{2\pi i m_am_b}{k}}\sum_{j=0,\frac{1}{2},\ldots}^{\frac{k-2}{2}}
e^{-\frac{2\pi i m_b}{k}(2j+1)} {\cal C}^{j}_{-2j-1-2m_a}(q,z)\,.
\end{align}

\item for the holomorphic part of a Liouville factor:
\begin{align}\label{chiholochar}
\tilde{\chi}_{hol;m_a,m_b}(-;l)
&=e^{\frac{2\pi im_am_b}{l}}\sum_{2j-1=0}^{l-1}e^{\frac{2 \pi i m_b}{l} (2j-1)} Ch(j;-\frac{1}{2}+m_a;q,z).
\end{align}

\item for the holomorphic part of a cigar factor:
\begin{align}\label{chicosholochar}
\tilde{\chi}_{hol;m_a.m_b}(-;l)^{\mathbb{Z}_l}
&=e^{\frac{2\pi im_am_b}{l}}\sum_{2j-1=0}^{l-1}e^{-\frac{2 \pi i m_b}{l} (2j-1)} Ch(j;-\frac{1}{2}-(2j-1)+m_a;q,z).
\end{align}
\end{itemize}
The completed twisted blocks for non-compact factors are recorded in appendix
\ref{characters}.

\section{Mirror symmetry for Gepner models}\label{mirrorsection}
In this section we recapitulate the construction of mirror Gepner
models \cite{Greene:1990ud},  generalized to include non-compact conformal field
theories. 

\subsection{Non-compact Gepner models}
Gepner's construction of string compactifications
in terms of exactly solvable $N=2$ superconformal field theories \cite{Gepner:1987qi}
can be
suitably 
extended to include factor models with central charge larger than $3$ (see e.g.
 \cite{Ashok:2007ui}).
We study non-compact Gepner models consisting of $p$ minimal models at
levels $k_i$ and $q$ non-compact models at levels $l_j$ tensored with $\mathbb{R}^{d-1,1}$.
They can be characterized in the light-cone as having
a 
\be
U(1)_2^{\frac{d-2}{2}} \times U(1)_2^{p+q} \times \prod_{i=1}^p U(1)_{k_i}
\times \prod_{j=1}^q U(1)_{l_j}
\ee
 worldsheet current algebra. The level $2$
factors refer to worldsheet fermion numbers, and the $U(1)$ current algebras
at level $k_i$ and $l_j$ are the R-currents of compact and non-compact
$N=2$ superconformal field theories. We have the corresponding charge
vectors $r$:
\be
r=(s_{-\frac{d-4}{2}},\dots,s_0,s_1,\dots,s_{p+q};
n_1, \dots, n_p; -2m_1, \dots , -2m_q),
\ee
 with inner product:
 \be r^{(1)} \cdot r^{(2)} = -\frac{
  s_{-\frac{d-4}{2}}^{(1)} s_{-\frac{d-4}{2}}^{(2)}}{4} \dots + \frac{
  n_{1}^{(1)} n_{2}^{(2)}}{2k_1} \dots - \frac{ 2m_{1}^{(1)}
  2m_{2}^{(2)}}{2l_1} \dots 
\label{scalprod}
\ee We introduce a vector $\beta_0$ such
that twice its inner product with the left-moving charge vector is
proportional to the left-moving R-charge. It satisfies $\beta_0\cdot \beta_0 = -1$. 
We fix conventions such that $\beta_0$ is equal to:
\be
\beta_0 = (1,\dots,1,1,\dots,1;1,\dots,1;1,\dots,1).
\ee
If we start from a model diagonal in the charge lattice quantum
numbers, then we must perform an orbifold to render the model local on
the worldsheet, in the sense of containing only purely NS or purely
Ramond states. The necessary $\mathbb{Z}_2^{p+q+(d-4)/2}$ orbifold
involves discrete torsion \cite{Greene:1990ud}. To obtain the type II Gepner model,
one further performs an integer R-charge orbifold, and a $\mathbb{Z}_2$ GSO
projection.

\subsection{Mirror symmetry through orbifolds}

Mirror symmetry is implemented in Gepner models through orbifolding by
a subgroup of the discrete group $G_{phase}=\prod_{i=1}^p
\mathbb{Z}_{k_i} \times \prod_{j=1}^q \mathbb{Z}_{l_j}$ of the
$U(1)_R$ symmetries of the factor theories \cite{Greene:1990ud}.  The
integer R-charge orbifold $\mathbb{Z}_n$ is already such a
subgroup\footnote{Here we ignore the fermionic entries in the charge
  vectors, which we can do if we allow for flat space charge
  conjugation.}.  The maximal subgroup $H$ of the group $G_{phase}$
which preserves space-time supersymmetry gives rise to the mirror
theory. Thus, the group $H$ will be the maximal subgroup of
$G_{phase}/\mathbb{Z}_n$ which preserves the condition that the left
and right R-charge remain integer.  Let us denote the original Gepner
model by $M_1$, and the mirror model by $M_2=M_1/H$.  Then if we
consider orbifolds of theory $M_1$ by the subgroup $H_1 \subset H$, we
will find that the theory $M_1/H_1$ is mirror to the theory
$M_2/(H/H_1)$.

We further note that the maximal allowed orbifold will give rise to
(the GSO projection of) the T-dual of the original model (before GSO).
In the T-dual, all left-moving angular momenta will have an opposite
sign.  These statements are true for the compact theory because a
$\mathbb{Z}_k$ orbifold of the minimal model gives rise to its
T-dual. For a singular non-compact theory (described by a purely
linear dilaton background), the statement also holds.
For the deformed or resolved non-compact theories, we need a
mild modification. For instance, the $\mathbb{Z}_l$ orbifold of
Liouville theory at radius $R=\sqrt{l \alpha'}$ is Liouville theory at
radius $R=\sqrt{\alpha'/l}$. That is T-dual to the {\em cigar} theory
at radius $R=\sqrt{l \alpha'}$. Thus orbifolding is equivalent to T-duality
only for the compact factors. For the non-compact factors, we must combine
orbifolding with an exchange of deformation and resolution in order to obtain
the T-dual model. 
We confirm this picture by the direct evaluation of elliptic genera for the mirror pair.

\subsection{Models with central charge $c=6$}
\label{c=6}

In this subsection, we get our feet wet with 
simple examples of
non-compact Gepner models with central charge $c=6$, and make some
preliminary observations. We concentrate on models involving one
compact and one non-compact model at equal levels.
As a starting point, we take a product of a minimal model with central charge
$c=3-6/k$ and an
$N=2$ Liouville theory at radius $R=\sqrt{k \alpha'}$ with central
charge $c=3+6/k$. The integer
R-charge orbifold is an orbifold by the
 group $\mathbb{Z}_k$. The conformal
field theory describes strings propagating on a space which
is asymptotically locally flat, with a linear dilaton slope.
It has a deformed $\mathbb{C}^2/\mathbb{Z}_k$ singularity at the
center. See \cite{Ashok:2007ui, Israel:2004ir} for  detailed
discussions. The elliptic genus of this theory is given by the orbifold
formula applied to the two factor theories:
\begin{align} 
\chi(k;k)^{\mathbb{Z}_k} &= \frac{1}{k}\sum_{m_a, m_b=0}^{k-1} 
\tilde{\chi}_{m_a,m_b}(k;-)\tilde{\chi}_{m_a,m_b}(-;k)\cr
&= \frac{1}{k} \frac{i \theta_{11}(q,z)}{\eta(q)^3} 
\sum_{m_a, m_b=0}^{k-1}\sum_{j=0, \frac{1}{2},\ldots}^{\frac{k-2}{2}}
e^{\frac{2\pi i m_b (2j+1)}{k}} 
q^{\frac{m_a^2}{k}} z^{\frac{2m_a}{k}}\, 
\cr &\qquad\qquad
{\cal C}^j_{2j+1-2m_a}(q,z) \, 
\hat{A}_{2k}(z^{\frac{1}{k}}q^{\frac{m_a}{k}}e^{\frac{2\pi i m_b}{k}}, z^2q^{2m_a} ;q)\,.
\label{M1c=6}
 \end{align}

\subsubsection{The ground states}
To link our results to known results on massless states, we observe that we can recuperate the Poincar\'e
polynomial of these models from these expressions, by taking the limit
that projects onto left-moving (R-charged weighted) Ramond ground
states. One thus 
recovers the results described for 
instance in \cite{Ashok:2007ui}.

\subsubsection{Mirror symmetry}
To advance our analysis of mirror symmetry we first observe that, for this $c=6$ model, the
maximal group $H$ of phase symmetries that we can mod out by while
preserving supersymmetry is trivial. Thus, the model must be
self-mirror. This can also be seen as a consequence of the
(generalized) hyperk\"ahler structure of the target space. We conclude that the 
elliptic genus of the model has to be equal to the elliptic genus of a diagonal
minimal model times the cigar model at radius $\sqrt{k \alpha'}$ modded out by
the integer R-charge and GSO projection. The latter elliptic genus is given by:
\begin{align}
\chi'(k;k)^{\mathbb{Z}_k} &=
 \frac{1}{k}\sum_{m_a, m_b=0}^{k-1} 
\tilde{\chi}_{m_a,m_b}(k;-)\tilde{\chi}_{m_a,m_b}(-;k)^{\mathbb{Z}_k}\cr
&= \frac{1}{k^2}\frac{i\theta_{11}(\tau, \a)}{\eta^3}\sum_{m_a, m_b=0}^{k-1}\sum_{j=0,\frac{1}{2}, \ldots}^{\frac{k-1}{2}}
e^{\frac{2\pi i m_b (2j+1)}{k}} 
q^{\frac{m_a^2}{k}}  z^{ \frac{2 m_a}{k}}\,\,  {\cal C}^j_{2j+1-2m_a}(q,z) \,  \cr
&\qquad\times\sum_{m_a',m_b'\in \mathbb{Z}_k}q^{-\frac{m_a^{'2}}{k}}e^{-\frac{2\pi i m'_an'_a}{k}}
\hat{A}_{2l}(z^{\frac{1}{k}}q^{\frac{m_a+m_a'}{k}}e^{\frac{2\pi i (m_b+m_b')}{k}},z^2q^{2m_a};q)\,.
\label{M2c=6}
\end{align}
The equality of the elliptic genera in equations (\ref{M1c=6}) and (\ref{M2c=6}) is non-trivial.
To prove the equality, it is useful to render the $N=2$ superconformal representation content
of the compact and non-compact elliptic genera manifest. 
In particular, let us write the holomorphic part of the elliptic genus 
\eqref{M1c=6} in terms of the characters of the minimal model and the
analogous extended characters 
\eqref{chiMMchar} and
\eqref{chiholochar}:
\begin{align}
\chi_{hol}(k;k)^{\mathbb{Z}_k} &= \sum_{j_1,j_2}\sum_{m_a,m_b \in \mathbb{Z}_k}e^{\frac{2\pi i m_b}{k}[(2j_1+1)+(2j_2-1)]}{\cal C}^{j_1}_{2j_1+1-2m_a}(q,z)Ch(j_2;-\frac{1}{2}+m_a;q,z) \ .
\end{align}
The sum over $m_b$ imposes the GSO constraint and relates the spin of
the two individual factors. In order to render the mirror interpretation manifest, we
shift the twisted sector label $m_a$  by $-(2j_2-1)\,$. We then use the
integer R-charge constraint in the angular momentum quantum number of the minimal
model character, to
end up with the final expression:
\begin{align}
\chi_{hol}(k;k)^{\mathbb{Z}_k} =& \sum_{j_1,j_2}\sum_{m_a,m_b \in \mathbb{Z}_k}e^{\frac{2\pi i m_b}{k}[(2j_1+1)+(2j_2-1)]}\cr
&\hspace{1in}{\cal C}^{j_1}_{-2j_1-1-2m_a}(q,z)Ch(j_2;-\frac{1}{2}+m_a -(2j_2-1);q,z) \,.
\end{align}
Repackaging this in terms of the twisted blocks, we find
\be \chi(k;k)^{\mathbb{Z}_k} = \frac{1}{k}\sum_{m_a, m_b=0}^{k-1}
\tilde{\chi}_{m_a,m_b}(k;-)^{\mathbb{Z}_k}\tilde{\chi}_{hol;m_a,m_b}(-;k)^{\mathbb{Z}_k}
\ee We recognize this to be the elliptic genus of an anti-diagonal
minimal model times the cigar theory at $R=\sqrt{k \alpha'}$, the
whole orbifolded by $\mathbb{Z}_k$. Performing a similar calculation
for the non-holomorphic long multiplet contributions gives rise to the
modular completion of the above formula.

In order to fully appreciate the relation between expressions
\eqref{M1c=6} and \eqref{M2c=6}, we have to go a bit further.  We will
give more details in the intricate $c=9$ examples, but we already
outline the idea here. We wish to re-interpret the mirror model as
an orbifold of a diagonal model. For this purpose, we note that the elliptic genus
of the anti-diagonal minimal model is related to the diagonal minimal
model elliptic genus through a sign flip in the second argument
$\alpha$, and the addition of an overall sign (see equation
(\ref{compflip})). Analogously the non-compact elliptic genus is
invariant under such a sign flip (see equation (\ref{noncompflip})).
Rewriting in terms of the twisted blocks we find
\begin{align}
\chi(k;k)^{\mathbb{Z}_k}(\tau,\alpha) &=
- \frac{1}{k}\sum_{m_a, m_b=0}^{k-1} 
\tilde{\chi}_{m_a,m_b}(k;-)(\tau,-\a)\tilde{\chi}_{m_a,m_b}(-;k)^{\mathbb{Z}_k}(\tau,\a)\cr
&=-\chi'(k;k)^{\mathbb{Z}_k}(\tau,-\alpha) \,.
\end{align}
Therefore the elliptic genus of the original model is self-mirror and
furthermore equal to the elliptic genus \eqref{M2c=6}, up to an
overall sign and a sign flip in the second argument $\alpha$. The
calculation provides a proof of a non-trivial relation between
products and sums of theta-functions and completed Appell-Lerch
sums. In the following sections, we will consider more involved
examples of mirror symmetry, including infinite families of mirror
pairs, and many more details on the long multiplet contributions, in
the context of non-compact Gepner models with central charge $c=9$.

\section{Models with  central charge $c=9$}
\label{c=9}
In this section we will study two types of models with central charge
$c=9$. The first type has tensor products of two minimal models and a
Liouville/cigar model. The second type has a single minimal model
tensored with two non-compact factors. 
We consider supersymmetric orbifolds of these models and exhibit
families of mirror pairs. 
 The first set serves to generate an
infinite set of mirror pairs (see also
\cite{Ashok:2007ui}), and illustrates in a fairly simple
setting how mirror symmetry acts on elliptic genera
 in the non-compact case. The second set analyzes more deeply how mirror
symmetry operates in the long multiplet sector.

\subsection{The $(2k,2k;k)$ model} \label{family} 

{From} the general discussion on mirror symmetry via orbifolds, it is clear that we must
identify the largest subgroup $H$ of the phase symmetries with which
one can orbifold and still preserve supersymmetry. Let us perform this
calculation for the $(2k,2k;k)$ model corresponding to two diagonal
minimal models and one $N=2$ Liouville theory at radius
$R=\sqrt{k\alpha'}$. There exists a Landau-Ginzburg (LG) model which
flows to this conformal field theory in the infrared 
and it is sometimes convenient to think in terms of such
a description. The LG model contains three chiral superfields
$X_1,X_2$ and $Y_3$ with superpotential:
\be
W=X_1^{2k}+X_2^{2k}+e^{-kY_3}\,.
\ee
The phase symmetries of the model are given by
$G_{phase}=\mathbb{Z}_{2k}\times\mathbb{Z}_{2k}\times \mathbb{Z}_k$.
We can identify the elements of the group $G_{phase}$ with charge
vectors in the following manner. A group element corresponds to a
charge vector $\gamma$ if it multiplies a state with diagonal charge
vector $r$ by the phase $e^{2 \pi i \gamma \cdot r}$. We can choose
generators $\gamma_i$ in each factor of the group $G_{phase}$ such that
$\gamma_i$ is the charge vector with entry $2$ in the spot corresponding
to the relevant $U(1)$ charge (see definition (\ref{scalprod})).

We identify the group $G = (\mathbb{Z}_{2k} \times
\mathbb{Z}_{2k} \times \mathbb{Z}_k)/\mathbb{Z}_{2k}$ as the subgroup
by which we can divide after taking into account the integer R-charge
orbifold $\mathbb{Z}_n=\mathbb{Z}_{2k}$.  The maximal subgroup $H$ of
$G$ that preserves supersymmetry corresponds to charge vectors
$\beta_m$ which are
 integer linear combinations
of the  charge vectors $\gamma_i$.
The generators $\beta_m$ need to satisfy:
\be
\beta_m = \sum_i c_m^i \gamma_i\quad \text{and} \quad  \beta_m \cdot \beta_0
\in \mathbb{Z}.
\ee
Using our conventions for $\beta_0$, this is equivalent to
\be
\sum_{i=1}^p \frac{c_m^i}{k_i} - \sum_{j=1}^q \frac{c^j_m}{l_j} \in \mathbb{Z}\,,
\ee  
where the $k_i$ are the levels of the minimal models while the $l_i$
refer to the levels of the non-compact models. The $c_m^i$ are
integers. In our specific
example, we have three coefficients $c_m^{i=1,2,3}$ for each
generator
$\beta_m$, which have to satisfy:
\begin{eqnarray}
+\frac{c_m^1}{2k}  +\frac{c_m^2}{2k} - \frac{c_m^3}{k} & \in & \mathbb{Z}.
\end{eqnarray}
The integers $c^i_m$ are defined modulo $(2k,2k,k)$. The integer R-charge
orbifold corresponds to $c^i_m = (1,1,1)$. 
Thus, we can use the gauging of the integer R-charge orbifold to put the
last entry to zero. Note also that if we consider the element of the R-charge
orbifold group that squares to one we find that it corresponds to the
vector $c^i_m=(+k,-k,0)$.
Thus, we conclude that the elements of the group $H$ have
representatives where the first two entries are opposite and that
these entries are only non-trivial modulo $k$. We therefore find that the
group $H$ is the $\mathbb{Z}_{k}$ group generated by multiplication
of the phases of $X_1$ and $X_2$ by $e^{\frac{2 \pi i}{2k}}$ and
$e^{-\frac{2 \pi i}{2k}}$ respectively.

\subsubsection{An infinite family of mirror pairs}

To generate an infinite family of mirror pairs, we can consider
subgroups of the group $H$. We suppose that the level $k$ of our
initial models is equal to the product of two positive integers $k=
k_{1} k_2$.  We can then consider orbifolds of our diagonal model with
the subgroup $\mathbb{Z}_{2k}\times \mathbb{Z}_{k_1}$ 
(or strictly
speaking, their semi-direct product) where the first factor
corresponds to the integer R-charge orbifold and the second factor to
the subgroup $\mathbb{Z}_{k_1}$ of the group $H$ generated by the
phase multiplication $e^{\pm 2 \pi i \frac{k_2}{2 k}}$ acting on the
fields $X_{1,2}$. 
Each group element of the orbifold group is labeled by a pair of integers $(m, n)$, taking values in $\mathbb{Z}_{k}$ and $\mathbb{Z}_{k_1}$ respectively.

\subsubsection*{Details of the calculation}

In what follows, 
we begin with the elliptic genus of this doubly orbifolded model and
show, analogous to what was done for the $c=6$ case, that we are able
to rewrite it as the elliptic genus of the mirror model.
In this case, the mirror is a $\mathbb{Z}_{2k} \times \mathbb{Z}_{k_2}$
orbifold of a product conformal field theory with two minimal model
factors and the cigar conformal field theory.

We start out with the holomorphic part of the orbifolded elliptic
genus written in terms of the twisted blocks. It depends only on
the charges of the fields under the orbifold group:
\begin{align}
\chi_{hol}(2k,2k;k)^{\mathbb{Z}_{2k},\mathbb{Z}_{k_1}}=&
 \frac{1}{2kk_1} \sum_{m_a,m_b \in \mathbb{Z}_{2k}}
\sum_{n_a,n_b \in \mathbb{Z}_{k_1}}
\tilde{ \chi}_{m_a+k_2n_a,m_b+k_2n_b} (2k;-)\cr
&\hspace{1.2in}\tilde{\chi}_{m_a-k_2n_a,m_b-k_2n_b} (2k;-)
\tilde{\chi}_{hol;m_a,m_b}(-;k).
\end{align}
The first minimal model factor contributes \be
\tilde{\chi}_{m_a+k_2n_a,m_b+k_2n_b}(2k;-) = e^{-\frac{2\pi i
    (m_a+k_2n_a)(m_b+k_2n_b)}{2k}}\sum_{j_1=0,\frac{1}{2},\ldots}^{\frac{2k-2}{2}}e^{\frac{2\pi
    i (m_b+k_2n_b)(2j_1+1)}{2k}} {\cal
  C}^{j_1}_{2j_1+1-2(m_a+k_2n_a)}(q,z)\,, 
\ee 
and similarly for the second minimal model factor with a sign flip for the $n_a, n_b$ quantum numbers. For the Liouville sector, we have
\be
\tilde{\chi}_{hol; m_a,m_b}(-;k) (\tau,\alpha) =e^{\frac{2\pi im_am_b}{k}}\sum_{2j_3-1=0}^{k-1}e^{\frac{(2j_3-1)}{k}2\pi i m_b} Ch(j_3;-\frac{1}{2}+m_a;q,z)\,.
\ee
Putting all the factors together we obtain:
\begin{align}
\chi_{hol}(2k,2k;k)^{\mathbb{Z}_{2k},\mathbb{Z}_{k_1}}
=& \frac{1}{2kk_1}\sum_{j_1, j_2, j_3}  \sum_{m_a,m_b \in \mathbb{Z}_{2k}} \sum_{n_a,n_b \in \mathbb{Z}_{k_1}}\cr
&  e^{\frac{2\pi i m_b}{2k}((2j_1+1) + (2j_2+1)+2(2j_3-1))}  \, e^{\frac{2\pi i n_b}{2k_1}((2j_1+1) - (2j_2+1)-2k_2n_a)} 
\cr
&   {\cal C}^{j_1}_{2j_1+1-2(m_a+k_2n_a)}(q,z) {\cal C}^{j_2}_{2j_2+1-2(m_a-k_2n_a)}(q,z)
Ch(j_3;-\frac{1}{2}+m_a;q,z)\,.\cr
\end{align}
The sum over the twist insertion $m_b$ then imposes the desired integer R-charge constraint:
\be
\frac{j_1+j_2+2j_3}{k}\in \mathbb{Z}.
\ee
A second constraint arises from the sum over the values of $n_b$: \be
\frac{j_1-j_2-k_2n_a}{k_1} \in \mathbb{Z}\,.  \ee Indeed, for any
projection beyond the initial integer R-charge projection, we will find a
constraint between spins and a new quantum number.  In order to rewrite this as the
mirror elliptic genus we find it useful to eliminate the twisted
quantum numbers $n_a$ in terms of the spin quantum numbers. 
 In order to solve for the second constraint, recall
that the angular momentum quantum number in the minimal model factor
is defined modulo twice the level. Using this we find
that there are precisely $k_2$ solutions to the second equation (where
$k=k_1\cdot k_2$). Solving for $n_a$, we substitute:
\be\label{Cnew}
k_2 n_a = j_1-j_2+n'_a k_1 \quad \text{with}\quad n'_a \in \mathbb{Z}_{k_2} \,.
\ee
This leads to 
\begin{align}
\chi_{hol}(2k,2k;k)^{\mathbb{Z}_{2k},\mathbb{Z}_{k_1}}
= & \frac{1}{2kk_2}\sum_{j_1, j_2, j_3}  \sum_{m_a,m_b \in \mathbb{Z}_{2k}}\sum_{n'_a, n'_b\in \mathbb{Z}_{k_2}}
\cr
 & e^{\frac{2\pi i m_b}{2k}((2j_1+1) + (2j_2+1)+2(2j_3-1))}  e^{\frac{2\pi i n'_b}{2k_2}((2j_1+1)-(2j_2+1)+2k_1 n'_a)}\cr
 & \sum_{n'_a\in \mathbb{Z}_{k_2}}{\cal C}^{j_1}_{2j_2+1-2(m_a+k_1n'_a)}(q,z){\cal C}^{j_2}_{2j_1+1-2(m_a-k_1n'_a)}(q,z)
Ch(j_3;-\frac{1}{2}+m_a;q,z)\,.\cr
\end{align}
The sum over the integer $n'_b \in \mathbb{Z}_{k_2}$
ensures that the numbers $n'_a$, $j_1$ and $j_2$ satisfy  the constraint \eqref{Cnew}. We now use
the integer R-charge constraint to eliminate the spin $j_2$ in the first
minimal model and the spin $j_1$ in the second minimal model character.
After a shift in the $m_a$ variable by $-(2j_3-1)$, we obtain our final expression:
\begin{align}
\chi_{hol}(2k,2k;k)^{\mathbb{Z}_{2k},\mathbb{Z}_{k_1}}
=& \frac{1}{2kk_2} \sum_{j_1, j_2, j_3}\sum_{m_a,m_b \in \mathbb{Z}_{2k}} e^{-\frac{2 \pi i m_b}{2k} [(2j_1+1) +(2j_2+1) + 2 (2j_3-1)]} \cr
&\sum_{n'_a, n'_b\in \mathbb{Z}_{k_2}} e^{\frac{2\pi i n'_b}{2k_2}((2j_1+1)-(2j_2+1)+2k_1 n'_a)}
 {\cal C}_{-2j_1 -1-2(m_a+k_1 n'_a)}^{j_1 } (q, z)\cr 
&{\cal C}_{-2j_2 -1-2 (m_a-k_1n'_a)}^{j_2 } (q, z)\,
Ch(j_3;-\frac{1}{2}+ m_a-2j_3-1; q, z)\,.
\end{align}
Rewriting this back in terms of the twisted blocks, we find that the final expression is equal to:
\begin{align}\label{2k2kkholo}
\chi_{hol}(2k,2k;k)^{\mathbb{Z}_{2k},\mathbb{Z}_{k_1}}=&
\frac{1}{2kk_2}\sum_{m_a,m_b\in \mathbb{Z}_{2k}}\sum_{n'_a,n'_b\in \mathbb{Z}_{k_2}} \tilde{\chi}_{m_a+k_1n'_a,m_b+k_1n'_b}(2k;-)^{\mathbb{Z}_{2k}} \cr
&\hspace{1in}\tilde{\chi}_{m_a-k_1n'_a,m_b-k_1n'_b}(2k;-)^{\mathbb{Z}_{2k}}
\tilde{\chi}_{hol;m_a,m_b}(-;k)^{\mathbb{Z}_{k}} .
\end{align}
%
To infer the mirror we have more work to do. Firstly we have to ensure that
the non-holomorphic part of the orbifold elliptic genus can also be
written such that it is the appropriate modular completion of the
above mock modular form. Secondly we need to have the orbifold of a diagonal model
in order to read of the mirror.

\subsubsection{The long multiplet sector}
\label{single}

In order to complete our matching of elliptic genera of the mirror
pairs, we also need to check the equality for the states in the long
multiplet sector. For simplicity, we restrict to the case in which the levels
satisfy $k_1=k$ and $ k_2=1$. The generalization to the other cases is
straightforward. 
The remainder term of the orbifold elliptic genus takes the form 
\begin{align}
\chi_{rem}(2k,2k;k)^{\mathbb{Z}_{2k},\mathbb{Z}_k}
&= \frac{1}{2 k^2}
\sum_{m_a,m_b \in \mathbb{Z}_{2k}}\sum_{n_a,n_b \in \mathbb{Z}_k}\chi_{m_a+n_a,m_b+n_b}(2k-;)\cr
&\hspace{2in}\chi_{m_a-n_a,m_b-n_b}(2k-;)\chi_{rem;m_a,m_b}(-;k)\,.\cr
\end{align}
In order to proceed we require the twisted blocks
that correspond to the non-holomorphic piece of the elliptic genus.
These are given in appendix \ref{characters}.  Using these blocks
along with the expressions for the minimal model elliptic genera, we
obtain
\begin{align}
\chi_{rem}(2k,2k;k)^{\mathbb{Z}_{2k},\mathbb{Z}_k}&=\frac{1}{2k^2}\sum_{j_1, j_2}  \sum_{m_a,m_b \in \mathbb{Z}_{2k}}
e^{\frac{2\pi i m_b}{2k}((2j_1+1) + (2j_2+1))} 
\sum_{n_a,n_b \in \mathbb{Z}_k}\, e^{\frac{2\pi i n_b}{2k}((2j_1+1) - (2j_2+1)-2n_a)} \cr
&\qquad\qquad\sum_{w,n \in \mathbb{Z}} e^{2 \pi i \frac{ n m_b}{ k}} z^{\frac{n-kw+2 m_a}{k}}\ 
{\cal C}^{j_1}_{2j_1+1-2(m_a+n_a)}(q,z)
{\cal C}^{j_2}_{2j_2+1-2(m_a-n_a)}(q,z)\cr
&\qquad\qquad (-1) \frac{1}{\pi} i \frac{\theta_{11} (\tau,\alpha)}{\eta^3} 
\int_{-\infty -i \epsilon}^{+\infty -i \epsilon}
\frac{ds}{2is+n+kw} q^{\frac{s^2}{k}+ \frac{(n-kw+ 2 m_a)^2}{4k}} \bar{q}^{\frac{s^2}{k}+ \frac{(n+kw)^2}{4k}}\,.
\end{align}
The calculation follows the same scheme as the previous one. The sum
over the integers $m_b$ and $n_b$ again imposes the desired
constraints:
\begin{align}
\frac{2j_1+1}{2k}+\frac{2j_2+1}{2k} +\frac{n}{k} &\in \mathbb{Z} \, ,\cr
\frac{j_1}{k}-\frac{j_2}{k}- \frac{n_a}{k}  &\in \mathbb{Z}\,. 
\end{align}
We eliminate the twisted quantum numbers $n_a$ in terms of the spins and obtain:
\begin{align}
\chi_{rem}(2k,2k;k)^{\mathbb{Z}_{2k},\mathbb{Z}_k}
&= \frac{1}{2k}\sum_{j_1, j_2}  \sum_{m_a,m_b \in \mathbb{Z}_{2k}}
e^{\frac{2\pi i m_b}{2k}((2j_1+1) + (2j_2+1))} \cr
&\qquad\sum_{w,n \in \mathbb{Z}} e^{2 \pi i \frac{ n m_b}{ k}}
z^{\frac{n-kw+2 m_a}{k}} 
{\cal C}^{j_1}_{2j_2+1-2m_a}(q,z)\, 
{\cal C}^{j_2}_{2j_1+1-2m_a}(q,z)\cr
&\qquad (-1) \frac{1}{\pi} i \frac{\theta_{11} (\tau,\alpha)}{\eta^3}
\int_{-\infty -i \epsilon}^{+\infty -i \epsilon}
\frac{ds}{2is+n+kw} q^{\frac{s^2}{k}+ \frac{(n-kw+ 2 m_a)^2}{4k}} \bar{q}^{\frac{s^2}{k}+ \frac{(n+kw)^2}{4k}}\, . \nonumber
\end{align}
We substitute the integer R-charge constraint  in the angular momentum variable of the two minimal models and shift the variable $m_a$ to $m_a-n+kw$ and find:
\begin{align}
\chi_{rem}(2k,2k;k)^{\mathbb{Z}_{2k},\mathbb{Z}_k}
&= \frac{1}{2k}\sum_{j_1, j_2}  \sum_{m_a,m_b \in \mathbb{Z}_{2k}}
e^{\frac{2\pi i m_b}{2k}((2j_1+1) + (2j_2+1))} \cr 
&\qquad\sum_{w,n \in \mathbb{Z}} e^{2 \pi i \frac{ n m_b}{ k}}
z^{\frac{-n+kw+2 m_a}{k}} 
{\cal C}^{j_1}_{-2j_1-1-2m_a}(q,z)\, 
{\cal C}^{j_2}_{-2j_2-1-2m_a}(q,z)\cr
&\qquad (-1) \frac{1}{\pi} i \frac{\theta_{11} (\tau,\alpha)}{\eta^3}
\int_{-\infty -i \epsilon}^{+\infty -i \epsilon}
\frac{ds}{2is+n+kw} q^{\frac{s^2}{k}+ \frac{(-n+kw+ 2 m_a)^2}{4k}} \bar{q}^{\frac{s^2}{k}+ \frac{(n+kw)^2}{4k}}\,. \cr
\end{align}
We then flip the sign of the variable $m_b$, and find that all
individual factors combined indeed agree with the twisted blocks of
the mirror model:
\be
\chi_{rem}(2k,2k;k)^{\mathbb{Z}_{2k},\mathbb{Z}_{k}}=
\frac{1}{2k}
\sum_{m_a,m_b\in \mathbb{Z}_{2k}}\tilde{\chi}_{m_a,m_b}(2k;-)^{\mathbb{Z}_{2k}}\tilde{\chi}_{m_a,m_b}(2k;-)^{\mathbb{Z}_{2k}}
\tilde{\chi}_{rem;m_a,m_b}(-;k)^{\mathbb{Z}_{k}} .
\ee
This is the modular completion of the mock modular form defined in
equation \eqref{2k2kkholo} for $k_1=k$ and $k_2=1$. We thus extended the proof
of the equality of elliptic genera of mirror symmetric models to the
long multiplet sector.

Finally, we can  rewrite the formula for the mirror
elliptic genus in terms of characters which 
are more easily read as being associated to a diagonal
spectrum. We find that the mirror can be written as:
\begin{align}
\chi(2k,2k;k)^{\mathbb{Z}_{2k},\mathbb{Z}_{k}} (\tau,\alpha) =& 
\frac{1}{2k}
\sum_{m_a,m_b\in \mathbb{Z}_{2k}}
\tilde{\chi}_{m_a,m_b}(2k;-)(\tau,-\alpha) \cr
&\hspace{.8in}\tilde{\chi}_{m_a,m_b}(2k;-)(\tau,-\alpha)
\tilde{\chi}_{m_a,m_b}(-;k)^{\mathbb{Z}_{k}}(\tau,-\alpha),
\end{align}
where we have flipped the sign of the summation variables. 
We have used two facts which we already encountered while discussing the self-mirror $c=6$ example. Firstly, that the anti-diagonal minimal model elliptic genera are equal to their diagonal model
counterpart, up to an overall sign change and a change in the sign of
the second argument (see equation (\ref{compflip})).  Secondly, that
the elliptic genus of a non-compact model is equal to itself under the
sign flip of the second argument (see equation (\ref{noncompflip})).
Note how a sign flip in the left-moving angular momentum comes down to 
T-duality for the compact factor, which is self-dual under T-duality. 
For the non-compact factor, the sign flip changes its nature from Liouville
theory to cigar model.

Our final expression is consistent with our expectations about the
mirror model.  The original model $M_1$ was a $(2k,2k;k)$ model with
Liouville deformation at radius $\sqrt{k \alpha'}$. The
$\mathbb{Z}_{2k} \times \mathbb{Z}_{k_1}$ orbifold of the model gives
rise to the mirror $M_2$ of this model which is a $(2k,2k;k)$ model at
radius $R=\sqrt{k \alpha'}$ modded out by $\mathbb{Z}_{2k} \times
\mathbb{Z}_{k_2}$.  If the original model $M_1$ is (Liouville)
deformed, then the mirror $M_2$ is expected to be (cigar) resolved,
which is indeed the case. We have thus exhibited an infinite family of
models, parameterized by a pair of integers $(k_1, k_2)$ that are
mirror to one another and for which
 the
elliptic genera match.

\subsection{The $(k;2k,2k)$ model}
\label{twononcompact}

We next consider the model with two non-compact factors and one
minimal model. The non-holomorphic sector of this model has
qualitatively different features from the models of 
subsection \ref{family}
since it involves the modular completion of a product of two mock
modular forms. The Landau-Ginzburg description of the model 
is given
by the superpotential
\be
W=X_1^{k}+e^{-2kY_2}+e^{-2k Y_3}\,.
\ee
We consider the orbifold by the group $\mathbb{Z}_{2k} \times \mathbb{Z}_k$ generated by:
\begin{align}
(X_1,  e^{-Y_2},  e^{-Y_3}) 
&\longrightarrow \left(e^{\frac{2\pi i}{k}}X_1, 
 e^{\frac{2\pi i}{2 k}} e^{-Y_2}, e^{\frac{2\pi i}{2 k}}  e^{-Y_3} \right) \cr
(X_1,   e^{-Y_2},   e^{-Y_3}) 
&\longrightarrow \left(X_1,  e^{\frac{2\pi i}{2 k}} e^{-Y_2}
,   e^{-\frac{2\pi i}{2 k}}  e^{-Y_3} \right) \,.
\end{align}
For simplicity we only focus on the orbifold by the full group $H = \mathbb{Z}_k$. 

\subsubsection{The short multiplet bound states}

Using the twisted blocks in equations \eqref{chiMMchar} and
\eqref{chiholochar}, the holomorphic part of the elliptic genus of the
double orbifold takes the form:
\begin{align}
\chi_{hol}(k;2k,2k)^{\mathbb{Z}_{2k},\mathbb{Z}_k}
&= \frac{1}{2k^2}\sum_{j_1, j_2, j_3}  \sum_{m_a,m_b \in \mathbb{Z}_{2k}}\sum_{n_a,n_b \in \mathbb{Z}_k}
 \cr
&
e^{\frac{2\pi i m_b}{2k}(2(2j_1+1) + (2j_2-1)+(2j_3-1))}
e^{\frac{2\pi i n_b}{2k}((2j_2-1) - (2j_3-1)+2n_a)} \cr
&
{\cal C}^{j_1}_{2j_1+1-2m_a}\, Ch(j_2; -\frac{1}{2}+(m_a+n_a); q, z)
Ch(j_3;-\frac{1}{2}+(m_a-n_a); q, z)\,.\cr
\end{align}
We find the constraints:
%
\be
\frac{2j_1+j_2+j_3}{k}\in \mathbb{Z}
\qquad \text{and}\qquad
\frac{j_2}{k}-\frac{j_3}{k}+ \frac{n_a}{k}  \in \mathbb{Z}\,. 
\ee
As before we will find it useful to eliminate the twisted quantum
numbers $n_a$ in terms of the spins, while retaining the integer
R-charge constraint as it is. We then substitute the R-charge
constraint in the angular momentum variable of the two Liouville
factors. Redefining the variable $m_a$ variable to
$m_a'=m_a-(2j_1-1)$, we finally obtain the expression:
\begin{multline}
\chi_{hol}(k;2k,2k)^{\mathbb{Z}_{2k},\mathbb{Z}_k}=
\frac{1}{2k}
 \sum_{m_a,m_b \in \mathbb{Z}_{2k}}
 \sum_{j_1, j_2, j_3}
e^{-2 \pi i m_b \frac{2j_1 +j_2 + j_3}{k} }
 {\cal C}_{-2j_1 -1-2 m_a}^{j_1 } (q, z)\cr
Ch(j_2;-\frac{1}{2}+ m_a-2j_2-1; q, z)\,
Ch(j_3;-\frac{1}{2}+ m_a-2j_3-1; q, z)\,.
\end{multline}
Repackaging this in terms of the twisted blocks, we find:
\be\label{k2k2kholo}
\chi(k;2k,2k)^{\mathbb{Z}_{2k},\mathbb{Z}_k}=
\frac{1}{2k} 
\sum_{m_a,m_b \in \mathbb{Z}_{2k}} 
\tilde{\chi}_{m_a,m_b}(k-;)^{\mathbb{Z}_k}
\tilde{\chi}_{hol;m_a,m_b}(-;2k)^{\mathbb{Z}_{2k}}
\tilde{\chi}_{hol;m_a,m_b}(-;2k)^{\mathbb{Z}_{2k}}\,.
\ee
As was done in the earlier examples we turn now to a calculation of
the non-holomorphic completion of the elliptic genus in order to read
off the mirror model. The non-holomorphic contribution for
this model is qualitatively different in nature and throws up new and
interesting points.

\subsubsection{The long multiplet scattering states}

Schematically, the fully modular elliptic genus of this orbifold model can be decomposed into a holomorphic and non-holomorphic piece as follows\footnote{We elaborate on this point in section \ref{mockmodular}.}:
\be\label{schematic}
\chi = \chi^1\, \chi^2\, \chi^3 = \chi^1_{hol} \left(\chi^2_{hol} \chi^3_{hol}+ \left[\chi^2_{hol}\chi^3_{rem} + \chi^2_{rem} \chi^3_{hol}+\chi^2_{rem}\chi^3_{rem}\right]\right)\,,
\ee
where we have suppressed the summation indices over the twisted blocks
of the orbifold. The terms in the square parenthesis are the
non-holomorphic completion for the product of two mock modular forms.
The mirror analysis of the first two terms in this completion parallel
the discussion in the previous subsections and we do not show the
details of the calculation since we obtain the expected result
parallel to the one obtained in equation \eqref{k2k2kholo}. The last
term is of a new type, and we consider it in detail below. Denoting it
by $T_3$, and reinstating the missing summation indices, let us use
the twisted blocks for the non-holomorphic sector and write it out in
full glory:
\begin{align}
T_3&=
\frac{1}{2k^2}
\sum_{m_a,m_b = 0}^{2k-1}\sum_{n_a,n_b=0}^{k-1}\tilde{\chi}_{m_a,m_b}(k;-)
\tilde{\chi}_{rem;m_a+n_a,m_b+n_b}(-;2k)\tilde{\chi}_{rem;m_a-n_a,m_b-n_b}(-;2k)\cr
&=\frac{1}{2k^2}
\left(\frac{i}{\pi}\frac{\theta_{11}}{\eta^3}\right)^2\sum_{j_1}\sum_{w_1,n_1}\sum_{w_2,n_2}\sum_{m_a,m_b}\sum_{n_a,n_b}e^{\frac{2\pi im_b((2j_1+1)-m_a)}{k}}{\cal C}^{j_1}_{2j_1+1-2m_a}\cr
&\times e^{\frac{2\pi i}{2k}((m_b+n_b)(n_1+(m_a+n_a))+(m_b-n_b)(n_2+(m_a-n_a)))}
\times 
z^{\frac{n_1-2kw_1+2(m_a+n_a)}{2k}+\frac{n_2-2kw_2+2(m_a-n_a)}{2k}}\cr
&\qquad\times\int \frac{ds_1}{2is_1+n_1+2kw_1} 
q^{\frac{s_1^2}{2k}+\frac{(n_1-2kw_1+2(m_a+n_a))^2}{8k}}
\bar{q}^{\frac{s_1^2}{2k}+\frac{(n_1+2kw_1)^2}{8k}}\cr
&\qquad\qquad\times\int \frac{ds_2}{2is_2+n_2+2kw_2} 
q^{\frac{s_2^2}{2k}+\frac{(n_2-2kw_2+2(m_a-n_a))^2}{8k}}
\bar{q}^{\frac{s_2^2}{2k}+\frac{(n_2+2kw_2)^2}{8k}}.
\end{align}
The phase factors give rise to the two constraints:
\begin{align}\label{GSOthismodel}
&\frac{n_1-2kw_1}{2k}+\frac{n_2-2kw_2}{2k}+\frac{2j_1+1}{k} \in \mathbb{Z} \, , \cr
& \frac{n_1-2kw_1}{2k}-\frac{n_2-2kw_2}{2k}+\frac{2n_a}{2k} \in \mathbb{Z}\,.
\end{align}
%
We have used the fact that $n_a$ is defined modulo $k$. As before we can solve for the variable $n_a$ using the 
second constraint:
\be
2n_a=(n_2-2kw_1)-(n_1-2kw_1)\,.
\ee 
The variable $n_a$ appears in  two different
combinations with the other variables in both of the non-compact
factors. Let us label those combinations $e_1$ and $e_2$, where
\begin{align}
e_1= (n_1-2kw_1) +2(m_a+n_a)\qquad\text{and}\qquad
e_2=(n_2-2kw_2)+2(m_a-n_a)\,.
\end{align}
Substituting for $n_a$, we see that the
 combinations
$e_1$ and $e_2$ become
\be
e_1= n_2-2kw_2+2m_a \qquad\text{and}\qquad e_2=n_1-2kw_1+2m_a\,.
\ee
We  use the integer R-charge constraint in equation
\eqref{GSOthismodel} 
to obtain:
\be
e_1 = -(n_1-2kw_1)+2m_a-2(2j_1+1)\qquad \text{and}\qquad e_2=-(n_2-2kw_2)+2m_a-2(2j_1+1) \,.
\ee
Shifting the variable $m_a$ by $(-2j_1-1)$ and substituting for the combinations $e_i$ in
the expression for $T_3$, we find the final form:
\begin{align}
  T_3
  &=
\frac{1}{2k}
\left(\frac{i}{\pi}\frac{\theta_{11}}{\eta^3}\right)^2\sum_{j_1}
\sum_{w_1,n_1}\sum_{w_2,n_2}\sum_{m_a,m_b}e^{\frac{2\pi
      im_b(2(2j_1+1)+n_1+n_2)}{2k}}\,
  z^{\frac{(-n_1+2kw_1+2m_a)}{2k}+\frac{(-n_2+2kw_2+2m_a)}{2k}}\cr
  &\hspace{1in}\times\int \frac{ds_1}{2is_1+n_1+2kw_1}
  q^{\frac{s_1^2}{2k}+\frac{(-n_1+2kw_1+2m_a)^2}{8k}}
\bar{q}^{\frac{s_1^2}{2k}+\frac{(n_1+2kw_1)^2}{8k}}\cr
  &\hspace{1.5in}\times\int \frac{ds_2}{2is_2+n_2+2kw_2}
  q^{\frac{s_2^2}{2k}+\frac{(-n_2+2kw_2+2m_a))^2}{8k}}
\bar{q}^{\frac{s_2^2}{2k}+\frac{(n_2+2kw_2)^2}{8k}}{\cal
    C}^{j_1}_{-2j_1-1-2m_a}\cr
  &=\frac{1}{2k}
\sum_{m_a,m_b}\tilde{\chi}_{hol;m_a,m_b}(k;-)^{\mathbb{Z}_k}\tilde{\chi}_{rem;m_a,m_b}(-;2k)^{\mathbb{Z}_{2k}}\tilde{\chi}_{rem;m_a,m_b}(-;2k)^{\mathbb{Z}_{2k}}
  \,.
\end{align}
The factors agree with the twisted blocks of the mirror model.  
Indeed, one can now combine all terms in equation \eqref{schematic}
and rewrite the full elliptic genus as the integer R-charge orbifold
of an anti-diagonal minimal model at level $k$, tensored with the two
cigar theories at level $2k$.  Thus, all terms in the elliptic genera
confirm the mirror symmetry of the models, including the long
multiplet contributions.
We can also rewrite this as the elliptic genus of a diagonal minimal
model combined with two cigars (up to an overall minus sign, and a
minus sign in the second argument of the elliptic genus).  Note how
our calculation again gives rise to non-trivial identities between the
orbifolded product of two modular completed Appell-Lerch sums $\hat{A}$.

Finally, let us stress that our method, ultimately based on T-duality,
will work for any number of products of minimal models and
Liouville/cigar theories and their orbifolds.
\section{Notes on mock modular forms}
\label{mockmodular}
In this section, we make various remarks on mock modular forms, a
field which is in full development in both mathematics (see
e.g. \cite{Zwegers, Zagier}) and physics (see
e.g. \cite{Semikhatov:2003uc, Gaiotto:2008cd, Manschot:2011dj, Alim:2010cf, Huang:2011qx}).
We propose that the embedding of the mathematics of mock modular forms in
our present conformal field theory perspective provides a fruitful
point of view.

\subsection{The shadow}
\label{ashadow}
As a prelude to our discussion, it will be useful to introduce the
concept of a shadow. It is sometimes convenient to make explicit the
dependence of the twisted partition function (which is a real Jacobi
form) on the anti-holomorphic parameter $\bar{\tau}$. Once the
partition function is known, this dependence can be read off from its
anti-holomorphic derivative which we refer to as the shadow
\cite{Zwegers, Zagier}.  For starters, let us explicitly compute the shadow \cite{Zwegers}
of the elliptic genus of $N=2$ Liouville theory at radius $R=\sqrt{l
  \alpha'}$ directly from the partition function
\cite{Troost:2010ud}\footnote{The shadow was also obtained in this
  fashion by Sameer Murthy.}.  The shadow is defined (up to
normalization and conjugation) as the anti-holomorphic derivative of
the real Jacobi form $\chi(-;l)$:
\begin{eqnarray}
\chi_{shad} (-;l) &=& 
\partial_{\bar{\tau}} \chi(-;l)
\nonumber \\
&=& - \frac{1}{4 \sqrt{l \tau_2}} \frac{\theta_{11} (\tau,\alpha)}{\eta^3}
\sum_{w,n \in \mathbb{Z}}
z^{\frac{n-lw}{l}} 
( n + lw)
q^{\frac{(n-lw)^2}{4l}} \bar{q}^{\frac{(n+lw)^2}{4l}}
\nonumber \\
&=& 
 - \frac{1}{2} \sqrt{\frac{{l}}{ {\tau_2}}} \frac{\theta_{11} (\tau,\alpha)}{\eta^3}
\sum_{m \in \mathbb{Z}_{2l}}
\Theta_{m,l} (q,z^{\frac{2}{l}})
\Theta^{\frac{3}{2}}_{m,l}(\bar{q}),
\label{shadowexpl}
\end{eqnarray}
where we used the definitions of the theta-functions of weight $1/2$ and 
$3/2$ at level $l$:
\begin{eqnarray}
\Theta_{m,l} (q,z) &=& \sum_{p \in \mathbb{Z}}
q^{l (p + \frac{m}{2l})^2} z^{l ( p + \frac{m}{2l})}
\nonumber \\
\Theta^{\frac{3}{2}}_{m,l}(q) &=& \sum_{p \in \mathbb{Z}}
(p + \frac{m}{2l}) q^{l (p + \frac{m}{2l})^2}.
\end{eqnarray}
The shadow is a sum of terms which are the product of a holomorphic
theta-function of weight $1/2$, and an anti-holomorphic theta-function of weight $3/2$.

\subsection{The product of mock modular forms}
Modular forms exhibit a ring structure. In particular, the
product of modular forms gives rise to another modular form. For mock
modular forms, the corresponding ring structure is not yet fully
understood. We therefore believe that it is interesting to observe
that if mock modular forms can be interpreted as the holomorphic parts
of the elliptic genera of conformal field theories, then their product
can be interpreted as the holomorphic part of the elliptic genus of
the tensor product conformal field theory (as in equation
(\ref{product})). Thus, the tensor product operation on conformal
field theories can give rise to a natural product of mock modular
forms, or to a suggestion of how to extend the definition of mock
modular forms to include these products. Clearly, the completions of
these products of mock modular forms will include products of mock
modular forms and remainder functions, as well as the product of
remainder functions. Indeed, imagine we have two real Jacobi forms
$\chi^{1,2}$, 
which are modular completions of  mock modular forms $\chi_{hol}^{1,2}$,
then their product will have a remainder term of a new type:
\begin{eqnarray}
\chi^1 \chi^2 &=& \chi^1_{hol} \chi^2_{hol} + (\chi^1_{hol} \chi^2_{rem} + \chi^1_{rem} \chi^2_{hol}
+ \chi^1_{rem} \chi^2_{rem}).
\end{eqnarray} 
 These 
sums of products of holomorphic and non-holomorphic pieces give rise to generalized shadows including the product of
remainder terms (consisting of properly weighted modular integrals of
theta-functions) and the shadows of individual non-compact elliptic
genera (for example as in equation (\ref{shadowexpl})).
\subsection{The orbifolds of completions of mock modular forms}
We gave an explicit example of an orbifold of such a product of
completed mock modular forms in subsection
\ref{twononcompact}. It is clear that our construction gives rise to a large
class of real Jacobi forms that is non-trivial. The corresponding mock
modular forms may contain multiple poles\footnote{The typical shadow
  however will be different from the shadow for the double pole case
  discussed in \cite{DMZ}, where it is the sum of a product of
  holomorphic and anti-holomorphic modular forms. We thank Sameer
  Murthy for a discussion on this point.}.  Beyond the orbifolds discussed
in this paper, we can imagine many different types of mock modular
forms and their completions that can arise in physical
contexts. Instead of performing R-charge orbifolds as we have done up
to now, we can extend the orbifold group much further.

For instance, we can consider symmetric product orbifold groups. It is
straightforward to write down the elliptic genus of a symmetric
product orbifold, using its Lagrangian description in terms of a sum
over coverings of the torus by the torus. The result is a new Jacobi
form obtained from the seed through Hecke operators. The Hamiltonian
interpretation of the resulting formula could prove interesting.
Moreover, we can introduce discrete torsion in more general abelian or
non-abelian orbifolds, further enlarging the class of expressions that
one can obtain on the physics side, providing more examples of what
could be called (generalized) mock modular forms.

Yet another class of theories that can be examined, are Landau-Ginzburg
theories with mixes of polynomial potentials, and exponentials. One
can compute their elliptic genus using free field techniques.  For the
polynomials, one uses the techniques of \cite{Witten:1993jg} while for
the exponentials, one uses the approach of \cite{Troost:2010ud}. This
could potentially open up a whole new realm of mock modular forms,
corresponding to elliptic genera of conformal field theories that may
not be exactly solvable but that can be described as infrared limits
of supersymmetric field theories.

\subsection{Uniqueness}
Since the mathematics of mock modular forms is not yet set in stone,
it is harder at the moment to prove the uniqueness of modular
completions of the largest class of mock modular forms (see however
 \cite{Zwegers, Zagier, DMZ} for interesting results in this direction). In particular,
the approach (used for compact models) of identifying polar parts and
using ellipticity and modularity to prove equality of elliptic genera
is not yet available for generic completed mock modular forms (though it may
apply to the case of a single non-compact factor examined in
subsection \ref{single}). Such a general
mathematical theory could give rise to the physical statement that the
long multiplet sector matching is
guaranteed by ellipticity and modularity. That would provide interesting
information on the asymptotics of these non-compact Gepner models from their
bound state spectrum, and vice versa.

\section*{Acknowledgements}
We would like to thank Atish Dabholkar, Jeff Harvey, Amir Kashani-Poor, Albrecht Klemm, Sameer Murthy and Thomas Wotschke for interesting discussions and useful correspondence. We thank the authors of \cite{DMZ} for making a preliminary version of their work available to us.  S.A would like to thank the Chennai Mathematical Institute for hospitality during the completion of this work.
Our research is partly funded by the grant ANR-09-BLAN-0157-02.

\appendix

\section{Characters}
\label{characters}
\subsection{Minimal model characters}
One way to define $N=2$ minimal model characters is implicitly:
\be
\sum_{n \in \mathbb{Z}_{2k}} {\cal C}_n^{j (s)} (\tau,\alpha) \Theta_{n,k}(\tau,-\frac{2 \alpha}{k})
= \chi^j(\tau,0) \Theta_{s,2} (\tau,-\alpha).
\ee
We used the theta-functions defined by the formula:
\begin{eqnarray}
\Theta_{n,k}(\tau,\alpha) &=& \sum_{m \in \mathbb{Z}}
e^{2 \pi i \tau k (m + \frac{n}{2k})^2} e^{2 \pi i \alpha k ( m + \frac{n}{2k})}.
\end{eqnarray}
The Ramond sector ground states correspond to states with R-charges
$\pm ( (2j+1)/k - 1/2)$.  The characters for representations built on
ground states are $ {\cal C}_{2j+1}^{j (+1)}$ and $ {\cal
  C}_{-2j-1}^{j (-1)}$.  We 
 note  that these two lists are
in fact identical when we use the equivalence relation 
$(j,n,s) \equiv
(k/2-j-1,n+k,s+2)$. {From} their implicit definition, we find the
character transformation rule:
\be
 {\cal C}_n^{j (s)}  (\tau,\alpha+m_a \tau+ m_b) = q^{-\frac{c}{6} m_a^2} z^{-\frac{c}{3} m_a} 
e^{2 \pi i (\frac{n}{k} - \frac{s}{2}) m_b}  {\cal C}_{n-2 m_a}^{j (s-2m_a)}  (\tau,\alpha).
\ee
We also need the twisted Ramond sector characters ${\cal C}^j_n$ which 
we define as:
\begin{eqnarray}
{ \cal C}^j_n &=& {\cal C}^{j (1)}_n-{\cal C}_n^{j (-1)}.
\end{eqnarray}
They satisfy the transformation rule:
\begin{eqnarray}\label{Cjelliptic}
 {\cal C}_n^{j}  (\tau,\alpha+m_a \tau+ m_b) 
&=& (-1)^{m_a+m_b} q^{-\frac{c}{6} m_a^2} z^{-\frac{c}{3} m_a} 
e^{2 
\pi i \frac{n}{k} m_b}
  {\cal C}_{n-2 m_a}^{j}  (\tau,\alpha),
\end{eqnarray}
as well as the equality:
\begin{eqnarray}
{\cal C}^j_{-n}(\tau,\alpha) &=& - {\cal C}^j_{n}(\tau,-\alpha).
\label{compflip}
\end{eqnarray}

\subsection{Minimal model twisted blocks}

In computing the minimal model twisted blocks, we assume that for an
individual model we have a partition function in which we sum over
left and right spins which satisfy $s=\bar{s}$ modulo 2. This is a
diagonal sum in terms of NS and R sectors. We then find for the
elliptic genus: 
\be\label{chiinchar} 
\chi(k;-) =\frac{\theta_{11} (q,
  z^\frac{k-1}{k} )}{\theta_{11} ( q, z^\frac{1}{k} )}=
\sum_{j=0,\frac{1}{2},\dots}^{\frac{k-2}{2}} {\cal
  C}_{2j+1}^{j}(q,z)\,.  
\ee
The twisted blocks are:
\begin{align}
\chi_{m_a,m_b}(k;-) &= 
e^{2 \pi i \frac{c}{6} m_a m_b}
e^{2 \pi i \frac{c}{6} (m_a^2 \tau+ 2 m_a \alpha)}
 \sum_{j=0,\frac{1}{2},\dots}^{\frac{k-2}{2}} 
 {\cal C}_{2j+1}^{j} (\tau, \alpha+ m_a \tau + m_b)
\cr
&= e^{2 \pi i \frac{c}{6} m_a m_b}
(-1)^{m_a+m_b}
 \sum_{j=0,\frac{1}{2},\dots}^{\frac{k-2}{2}} 
e^{2 \pi i m_b \frac{2j + 1}{k} }
 {\cal C}_{2j+1-2m_a}^{j } (\tau, \alpha).
\end{align}
%
Inserting the standard phase $\epsilon$, we obtain
\begin{align}\label{chimambtakeone}
\tilde{\chi}_{m_a,m_b} (k;-)
&= e^{-\frac{2\pi i m_am_b}{k}} \sum_{j=0,\frac{1}{2},\dots}^{\frac{k-2}{2}} 
e^{2 \pi i m_b \frac{2j + 1}{k} }
 {\cal C}_{2j+1-2m_a}^{j } (\tau, \alpha).
\end{align}
We have used the known elliptic properties of the Ramond sector
characters in order to derive the twisted blocks.
 Equivalently, we can
perform the calculation using the ellipticity properties of the
theta-function.
We obtain
\be
\tilde{\chi}_{m_a,m_b} (k;-)=
z^{-\frac{m_a}{k}}
\frac{\theta_{11} (z^{(1-\frac{1}{k})}q^{-\frac{m_a}{k}} e^{-\frac{2\pi i m_b}{k}}) } 
{\theta_{11} (z^{\frac{1}{k}}q^{\frac{m_a}{k}} e^{\frac{2\pi i m_b}{k}} )}\,.
\ee
We note in passing that with this choice of phase $\epsilon$, the twisted
blocks of \cite{Kawai:1993jk} and \cite{Berglund:1993fj} agree. It 
remains to compare this to the sum of the Ramond sector characters.
We rewrite:
\begin{align}
\tilde{\chi}_{m_a,m_b}(k;-) &=z^{-\frac{m_a}{k}}\frac{\theta_{11}( z'^{(1-\frac{1}{k})}\, q^{-m_a}; q) }
{\theta_{11}(z'^{\frac{1}{k}}; q)}\,,
\end{align}
with
\be
z'=z q^{m_a}\, e^{2\pi i m_b}\,.
\ee
Using the elliptic property of the theta-function, we can write this as 
\begin{align}
\tilde{\chi}_{m_a,m_b}(k;-) 
&= (-1)^{m_a}q^{\frac{m_a^2}{2}(1-\frac{2}{k})}z^{m_a(1-\frac{2}{k})}e^{-\frac{2\pi i m_am_b}{k}}\frac{\theta_{11}( z'^{(1-\frac{1}{k})};q)}{\theta_{11}(z'^{\frac{1}{k}}; q)}\,.
\end{align}
The  ratio of theta functions can be expanded in terms of the
Ramond-sector characters as in equation \eqref{chiinchar}. 
We then again use the elliptic properties of the minimal model characters \eqref{Cjelliptic} to find
that the result agrees with equation \eqref{chimambtakeone}.
 We have come full circle.

\subsection{The $\mathbb{Z}_k$ orbifold and mirror symmetry}

Consider the $\mathbb{Z}_k$ orbifold the $N=2$ minimal model (with
$s=\bar{s}$ mod 2) of central charge $c=3-6/k$. Let us
calculate the elliptic genus of the orbifold:
\begin{align}
\chi(k;-)^{\mathbb{Z}_k} &= \frac{1}{k} \sum_{m_a,m_b \in \mathbb{Z}_k}
\tilde{\chi}_{m_a,m_b}(k;-)\cr
&= \frac{1}{k} \sum_{m,n \in \mathbb{Z}_k}   e^{-2 \pi i \frac{m_a m_b}{k}}
 \sum_{j=0,\frac{1}{2},\dots}^{\frac{k-2}{2}} 
e^{2 \pi i m_b \frac{2j +1}{k}}
 {\cal C}_{2j+1-2m_a}^{j } (\tau, \alpha).
\end{align}
The sum over the variable $m_b$ puts $m_a=2j + 1 \, (\mbox{mod} \, k)$ and adds a factor of
$k$. We can most easily eliminate $m_a$ from the sum and find:
\begin{align}
\chi(k;-)^{\mathbb{Z}_k} &=  \sum_{j=0,\frac{1}{2},\dots}^{\frac{k-2}{2}}  
{\cal C}_{-2j - 1}^{j }(\tau, \alpha) \nonumber \\
&= - \sum_{j=0,\frac{1}{2},\dots}^{\frac{k-2}{2}}  
{\cal C}_{2j+1}^{j }(\tau, -\alpha). 
\end{align}
This is one of the simplest examples of mirror symmetry in conformal
field theory. We recognize the previous to last line as the elliptic
genus of the anti-diagonal minimal model. 
Note that for these compact models,
the $\mathbb{Z}_k$ orbifold is equivalent to performing T-duality.

\subsection{Characters at $c>3$}
The elliptic genus of $N=2$ Liouville theory at radius $R=\sqrt{l \alpha'}$ 
contains a holomorphic part and a remainder term, namely it is a completed
Appell-Lerch sum $\hat{A}_{2l}$:
\begin{eqnarray}
\chi(;l) &=& \chi_{hol} + \chi_{rem}
\nonumber \\
&=& i \frac{\theta_{11} (\tau,\alpha)}{\eta^3} \hat{A}_{2l}(z^{\frac{1}{l}},
z^2; q)
\nonumber \\
\chi_{hol}(;l) &=&
i \frac{\theta_{11} (\tau,\alpha)}{\eta^3}
\sum_{m \in \mathbb{Z}} \frac{q^{lm^2} z^{2m}}{1-z^{\frac{1}{l}} q^m}
\nonumber \\
\chi_{rem} (;l) &=& 
- \frac{1}{\pi} i \frac{\theta_{11} (\tau,\alpha)}{\eta^3}
\sum_{w,n \in \mathbb{Z}}
z^{\frac{n-lw}{l}} 
\int_{-\infty -i \epsilon}^{+\infty -i \epsilon}
 \frac{ds}{2is+n+lw} q^{\frac{s^2}{l}+ \frac{(n-lw)^2}{4l}} \bar{q}^{\frac{s^2}{l}+ \frac{(n+lw)^2}{4l}}
\end{eqnarray} 
The holomorphic part of the Liouville elliptic genus can be expanded
in terms of the twisted Ramond sector characters. We have the
equation:
\begin{align}
\chi_{hol}(-;l)
&= 
\frac{i \theta_{11}(q,z)}{\eta^3}
\sum_{m \in \mathbb{Z}} \frac{q^{l m^2} z^{2m}}{1-z q^{lm}}
\sum_{2j-1=0}^{l-1} z^{\frac{2j-1}{l}} q^{m(2j-1)}\cr
&=  \sum_{2j-1=0}^{l-1} Ch (j;-\frac{1}{2};q,z)\,.
\end{align}
We notice that the elliptic genus is expressed as a sum over extended
characters. These correspond to ordinary characters summed over
spectral flow orbits that shift the angular momentum quantum number by
multiples of the level $l$.


The holomorphic part of the cigar elliptic genus can also be written
in terms of these extended characters:
\begin{align}
\chi_{hol}(-;l)^{\mathbb{Z}_k}(q,z)
&=\sum_{m=0,1,\ldots l-1}\sum_{w}\frac{i\theta_{11}(q,z)}{\eta^3}
\frac{q^{lw^2-mw}z^{2w-\frac{m}{l}} }{1-zq^{lw-m}}.
%
\cr
&=\sum_{2j-1=0}^{k-1}Ch(j;-\frac{1}{2}-(2j-1);q,z).
\end{align}
 The modular and ellipticity properties of the extended characters
are (for $m_a, m_b \in
\mathbb{Z}$): 
\be\label{chidzq} Ch (j;r';q,z
q^{m_a}e^{2\pi i m_b})= (-1)^{m_a+m_b}q^{- \frac{c}{6} m_a^2} z^{-
  \frac{c}{3} m_a} e^{\frac{2\pi i m_b(2j+2r')}{l}}Ch
(j;r'+m_a;q,z ).  \ee
The angular momentum of the representations corresponding
to these characters 
is $2j+2r'$.
We also have the following transformation rules for the holomorphic and 
remainder term of the elliptic genus:
\begin{eqnarray}
\chi_{hol}(-;l) (\tau,-\alpha)
 &=& \chi_{hol}(-;l) (\tau,\alpha)
 - \frac{i \theta_{11}}{\eta^3}     \sum_{m \in \mathbb{Z}}  q^{k m^2} z^{2m}
\nonumber \\
\chi_{rem}(-;l) (\tau,-\alpha) &=&  \chi_{rem}(-;l) (\tau, \alpha)
 + \frac{i \theta_{11}}{\eta^3}   \sum_{m \in \mathbb{Z}} q^{km^2} z^{2m}.
\end{eqnarray}
The extra term is a reminder of the ambiguity of the holomorphic part of the elliptic genus,
due to the bound state spectrum touching the delta-function normalizable continuum.
Together, these equations give rise to the equality:
\begin{eqnarray}
 \chi(;l) (\tau,-\alpha) &=& \chi(;l) (\tau, \alpha),
\label{noncompflip}
\end{eqnarray}
which can also be derived directly from the integral representation
of the non-compact elliptic genus.

\subsection{Twisted building blocks at $c>3$}

\subsubsection{Character formulae}
Using these properties, we can calculate the holomorphic part of the 
twisted blocks for the
Liouville and cigar elliptic genera:
\begin{align}
\chi_{m_a,m_b}(-;l) (\tau,\alpha) &=
e^{2 \pi i \frac{c}{6} m_am_b}
e^{2 \pi i \frac{c}{6} (m_a^2 \tau+ 2 m_a \alpha)}
 \sum_{2j-1=0}^{l-1} 
Ch(j;-\frac{1}{2};\tau, \alpha+ m_a \tau + m_b)\cr
&= (-1)^{m_a+m_b}e^{2 \pi i \frac{c}{6} m_am_b}
 \sum_{2j-1=0}^{l-1} e^{2 \pi i m_b \frac{2j-1}{l}} 
Ch(j;-\frac{1}{2}+m_a;\tau, \alpha).
\end{align}
We use the value of the central charge $c=3+6/l$, multiply by the
phase factor $\epsilon$ and obtain:
\begin{align}
\tilde{\chi}_{hol;m_a,m_b}(-;l)
&=e^{\frac{2\pi im_am_b}{l}}\sum_{2j-1=0}^{l-1}e^{\frac{(2j-1)}{k}2\pi i m_b} Ch(j;-\frac{1}{2}+m_a;q,z).
\end{align}
For the cigar, we find:
\begin{align}
\tilde{\chi}_{hol;m_a.m_b}(-;l)^{\mathbb{Z}_l}
&=e^{\frac{2\pi im_am_b}{l}}\sum_{2j-1=0}^{l-1}e^{\frac{-(2j-1)}{l}2\pi i m_b} Ch(j;-\frac{1}{2}-(2j-1)+m_a;q,z).
\end{align}

\subsubsection{Twisted blocks for the non-holomorphic sector}

For the continuous character part of the elliptic genus we find, for the Liouville theory
twisted block:
%
\begin{align}
\tilde{\chi}_{rem; m_a, m_b}(-;l) &=
(-1) \frac{1}{\pi} i \frac{\theta_{11} (\tau,\alpha)}{\eta^3}
\sum_{w,n \in \mathbb{Z}} e^{2 \pi i \frac{ (n+m_a) m_b}{l}}
z^{\frac{n-lw+2 m_a}{l}}\cr
&  \hspace{2cm}\int_{-\infty -i \epsilon}^{+\infty -i \epsilon}
\frac{ds}{2is+n+lw} q^{\frac{s^2}{l}+ \frac{(n-lw+ 2 m_a)^2}{4l}} \bar{q}^{\frac{s^2}{l}+ \frac{(n+lw)^2}{4l}}\,.
\end{align} 
For the cigar theory,  we end up with:
\begin{align}
\tilde{\chi}_{rem;m_a,m_b}(-;l)^{\mathbb{Z}_l} 
 &=
e^{2 \pi i \frac{ m_a m_b}{l}} 
(-1) \frac{1}{\pi} i \frac{\theta_{11} (\tau,\alpha)}{\eta^3}
\sum_{w,n \in \mathbb{Z}}
e^{-2 \pi i m_b \frac{n}{l}}
z^{- \frac{n-lw}{l} + \frac{2m_a}{l}}
\cr
&  \hspace{2cm}\int_{-\infty -i \epsilon}^{+\infty -i \epsilon}
\frac{ds}{2is+n+lw} q^{\frac{s^2}{l}+ \frac{(n-lw- 2 m_a)^2}{4l}} \bar{q}^{\frac{s^2}{l}+ \frac{(n+lw)^2}{4l}}\,.
\end{align}

\subsubsection{Exact expressions for twisted blocks}

Finally, we give the expressions for the complete twisted building blocks, for the Liouville theory:
\begin{align}
\tilde{\chi}_{m_a,m_b}(-;l)
&=  e^{\frac{2\pi i m_am_b}{l}}q^{\frac{m_a^2}{l}} z^{ \frac{2 m_a}{l}}\, 
\frac{i \theta_{11} ( \tau, \alpha)}{\eta^3}
\hat{A}_{2l} (z^{\frac{1}{l}} q^{\frac{m_a}{l}} e^{\frac{2 \pi i m_b}{l}} ,z^2 q^{2 m_a}  ; q),
\end{align}
and for the cigar theory at radius $R=\sqrt{l \alpha'}$: 
\begin{multline}
\tilde{\chi}_{m_a,m_b}(-;l)^{\mathbb{Z}_l}=  e^{\frac{2\pi i m_am_b}{l}}q^{\frac{m_a^2}{l}}  z^{ \frac{2 m_a}{l}}\, \frac{i\theta_{11}(\tau, \a)}{\eta^3}\cr
\times\frac{1}{l}\sum_{m_a',m_b'\in \mathbb{Z}_l}q^{-\frac{m_a^{'2}}{l}}e^{-\frac{2\pi i m'_an'_a}{l}}
\hat{A}_{2l}(z^{\frac{1}{l}}q^{\frac{m_a+m_a'}{l}}e^{\frac{2\pi i (m_b+m_b')}{l}},z^2q^{2m_a};q).
\end{multline}

\end{document}